\documentstyle[preprint,revtex,eqsecnum]{aps}
\def\noalignand{
\noalign{\vbox{\vskip\abovedisplayskip \hbox{and} \vskip\belowdisplayskip}}}
\begin{document}
\draft
\preprint{TRI-PP-92-125}
\preprint{December 1992}
\setlength{\baselineskip}{5ex}



\begin{title}
\begin{center}
{{\bf The electroweak theory of SU(3) $\times$ U(1)}}
\end{center}
\end{title}

\vspace{0.3cm}
\author{Daniel Ng}
\begin{instit}
\begin{center}
{\em TRIUMF, 4004 Wesbrook Mall, Vancouver, B.C., Canada V6T 2A3}
\end{center}
\end{instit}

\begin{center}
{\bf ABSTRACT}
\end{center}

An electroweak model of SU(3) $\times$ U(1) gauge group is studied.
{}From the group theoretical constraint, the symmetry breaking of
this model to the standard model occurs at 1.7~TeV or lower.
Hence the mass of the new neutral gauge boson is less than 1.7 TeV.
The $Y^\pm$ and $Y^{\pm\pm}$ masses are found to be less than
half of the $Z_2$ mass.
Thus, the decays $Z_2 \to Y^{++} Y^{--}$ with $Y^{++} \to 2\ell^+ \: (\ell
\!=\! e,\mu,\tau)$ is allowed, providing spectacular signatures at
future colliders.
{}From the flavor-changing neutral current processes, the representations of
quarks can be uniquely determined.  The neutrino-isoscalar scattering
experiments are also considered.

\vspace{2.0cm}
\newpage

\section{INTRODUCTION}

A model of SU(15), which includes doubly charged gauge bosons $(Y^{\pm\pm})$
and their isospin partners $(Y^\pm)$, was proposed by Frampton and Lee
\cite{fram&lee} two
years ago.  The model conserves baryon number in gauge interactions, thus
proton decay is naturally suppressed \cite{fram&keph}.
The process $e^-e^- \to \mu^-\mu^-$
would be the best experiment testing the existence of a doubly charge gauged
boson.  However, the only machine relevant to this process is operated at the
center-of-mass energy 1.112~GeV \cite{barber&etal}.
Nevertheless, there are $u$-channel
contributions to the Bhabha scattering $e^+e^- \to e^+e^-$ leading to the mass
lower bound $M_{Y^{++}} > 210$~GeV (95\% C.L.) \cite{fram&ng}
and right-handed current
contributions to muon decay leading to the bound $M_{Y^+} > 270$~GeV
(90\% C.L.) \cite{belt&etal}.

Motivated by a doubly charged gauge boson, a model of SU(3)$_L \times {\rm
U}(1)_X$ is introduced by Frampton \cite{frampton}
and Pisano {\em et al.} \cite{pis&ple}.
The former author looked for a simple solution which included
dileptons $Y^{\pm\pm}$; the latter
author argued that $Y^{--}$ is necessary in order to avoid unitarity violation
for $e^-e^- \to W^- Y^-$ at high energies.  $Y^{\pm\pm}$ and $Y^{\pm}$
are called dileptons because they couple to two leptons, thus they have
two units of lepton number.  Many other electroweak models
\cite{singer&etal} of
SU(3) $\times$ U(1) were suggested some years ago with different choices of
particle content.   Here, this model has miminal particle content
yielding some interesting new physics, such as stringent constraints on
the new gauge boson masses.

The anomaly in
this model is not cancelled within each generation.  However, the
representation of one of the quark generations is chosen in such a way
that the anomaly is cancelled among 3 generations.
Thus the number of generations is a multiple of 3.
Note that the third and the first generation quark multiplets
were chosen arbitrarily by the authors in Refs.\cite{frampton} and
\cite{pis&ple}, respectively.  In this paper, we
show that only the former choice provides a consistent phenomenology.

In most extended models such as E$_6$ and left-right symmetric (L-R) models,
the symmetry breaking scales can be as high as the grand unification scale.
Here, the breaking of SU(3)$_L \times {\rm U}(1)_X$
occurs at 1.7~TeV or less (but greater than 250~GeV).
Therefore, this model will be discovered or ruled out at the future
colliders.  We will organize this paper as follows:  Section~II describes the
model; in Secs.~III and IV, gauge boson and fermion masses are discussed.
Section~V investigates flavor-changing neutral current processes;
neutrino-hadron scattering is studied in Sec.~VI; finally, the conclusions are
presented in Sec.~VII.

\section{DESCRIPTION OF THE MODEL}

The simplest anomaly-free solution \cite{frampton},
which includes the standard model,
of a gauge symmetry SU(3)$_c \times {\rm SU}(3)_L \times {\rm U}(1)_X$
is given as follows:
\begin{mathletters}
\begin{eqnarray}
\psi_{1,2,3} = \pmatrix {e\cr \nu_e\cr e^c\cr} \: , \pmatrix {\mu \cr \nu_\mu
   \cr \mu^c \cr} \: , \pmatrix {\tau \cr \nu_\tau \cr \tau^c \cr}
     \qquad &\mathbin:& \qquad (1, \: 3^\ast, \: 0) \ , \\
Q_{1,2} = \pmatrix {u\cr d\cr D\cr} \: , \pmatrix {c\cr s\cr S\cr}
     \qquad &\mathbin:& \qquad (3, \: 3, \: -\textstyle\frac{1}{3} ) \ , \\
Q_3 = \pmatrix {b\cr t\cr T\cr} \qquad &\mathbin:& \qquad (3, \: 3^\ast, \:
     \textstyle\frac{2}{3} ) \ ,  \\
d^c, \: s^c, \: b^c \qquad &\mathbin:& \qquad \ \textstyle\frac{1}{3} \ ,  \\
u^c, \: c^c, \: t^c \qquad &\mathbin:& \qquad -\textstyle\frac{2}{3} \ , \\
D^c, \: S^c \qquad &\mathbin:&  \qquad \ \textstyle\frac{4}{3} \ ,  \\
T^c \qquad &\mathbin:& \qquad  \textstyle -\frac{5}{3} \ .
\end{eqnarray}
\end{mathletters}
where $D,\,S,\,T$ are new quarks.  For a minimal particle content, the anomaly
is not cancelled within each generation, but cancelled among three
generations by choosing the third quark generation as an SU(3)$_{\rm L}$
anti-triplet. So far, the choice of the quark generation
is arbitrary.  However, as we shall discuss later,
the third generation is chosen in order to have consistent phenomenology.

SU(3)$_L \times {\rm U}(1)_X$ will first be broken down to the standard model
SU(2)$_L \times {\rm U}(1)_Y$ by a nonzero vacuum expectation value (VEV) of a
triplet scalar $\langle \Phi \rangle^T = (0,\,0,\,u/\sqrt{2})$, yielding a
massive neutral gauge boson $(Z'$) and two charged gauge bosons $(Y^+,Y^{++})$
as well as new quarks $(D,\,S,\,T)$.  The breaking of SU(2)$_L \times {\rm
U}(1)_Y$ to U(1)$_{\rm em}$ can be achieved by $\langle \Delta \rangle^T =
(0,\,v/\sqrt{2}, \, 0)$ and $\langle \Delta ' \rangle^T = (v'/\sqrt{2},\, 0,\,
0)$.  In order to obtain acceptable masses for charged leptons, a sextet $\eta$
is necessary.  Hence, the required scalar multiplets are summarized as follows:
\begin{mathletters}
\begin{eqnarray}
\Phi = \pmatrix{\phi^{++}\cr \phi^+\cr \phi^0\cr} \qquad \qquad
         &\mathbin:& \qquad (1,\,3,\,1) \ ,  \\
\Delta = \pmatrix{\Delta_1^+\cr \Delta^0\cr \Delta_2^-\cr} \qquad \qquad
          &\mathbin:& \qquad (1,\,3,\,0) \ , \\
\Delta ' = \pmatrix{\Delta '^0\cr \Delta '^-\cr \Delta '^{--}\cr} \qquad
        \quad &\mathbin:& \qquad (1,\,3,\,-1) \ ,  \\
\noalignand
\eta = \pmatrix {\eta_1^{++} & \eta_1^+/\sqrt{2} & \eta^0/\sqrt{2} \cr
                 \eta_1^+/\sqrt{2} & \eta '^0 & \eta_2^-/\sqrt{2} \cr
                 \eta^0/\sqrt{2} & \eta_2^-/\sqrt{2} & \eta_2^{--} \cr}
      \qquad &\mathbin:& \qquad (1,\,6,\,0)
\ .
\end{eqnarray}
\end{mathletters}

\vspace{0.3cm}

\section{GAUGE BOSON MASSES}

To obtain the gauge interactions, let us first define the covariant derivative
for triplets
\begin{eqnarray}
D_\mu = \partial_\mu - ig\: \frac{\lambda^a}{2} \: W^a - ig_X \, X \:
\frac{\lambda^9}{2} \: V \ ,
\end{eqnarray}
where $\lambda^a \; (a\!=\!1,\cdots ,\,8)$ are the SU(3)$_L$ generators, and
$\lambda^9 = \sqrt{\frac{2}{3}}$~diagonal(1,1,1) are defined such that ${\rm
Tr}(\lambda^a \lambda^b) = 2\delta^{ab}$ and
 ${\rm Tr}(\lambda^9\lambda^9) = 2$.
$g$ and $g_X$ are the gauge coupling constants for SU(3)$_L$ and U(1)$_X$ with
their gauge bosons $W^a$ and $V$, respectively.  The covariant derivative for
the sextet is
\begin{eqnarray}
D_\mu \, \eta^{\alpha\beta} = \partial_\mu \, \eta^{\alpha\beta}
   - i\frac{g}{2}\, W^a
      \left[  {\lambda^a} ^\beta _ {\beta '} \, \eta^{\alpha\beta '}
    + {\lambda^a} ^\alpha _ {\alpha '} \, \eta^{\alpha '\beta} \right] \ .
\end{eqnarray}

As the triplet scalar $\Phi$ acquires a VEV, the symmetry SU(3)$_L \times {\rm
U}(1)_X$ breaks down to SU(2)$_L \times {\rm U}(1)_Y$, where $Y \equiv \sqrt{3}
(\lambda^8 + \sqrt{2} \, X \, \lambda^9)$ is the hypercharge.  The coupling
constant of U(1)$_Y$, $g'$ is given by
\begin{eqnarray}
\frac{1}{g'^2} = 3 \left( \frac{1}{g^2} + \frac{2}{g_X^2} \right) \ .
\end{eqnarray}
Therefore we obtain
\begin{eqnarray}
\frac{g_X^2}{g^2} = \frac{6 \, \sin^2 \theta_W}{1 \!-\!4 \: \sin^2 \theta_W} \
{}.
\end{eqnarray}
where $g'/g = \tan \theta_W$.  Therefore, $\sin^2 \theta_W$ has to be smaller
than 1/4 at the breaking scale.  Below this breaking scale, there are three
doublets, $(\Delta_1^+,\Delta^0)$, $(\Delta'^0,\Delta'^-)$ and
$(\eta^0,\eta_2^-)$,  one triplet
$(\eta_1^{++},\eta_1^+,\eta'^0)$, and three singlets $\eta_2^{--}$,
$\Delta_2^-$ and $\Delta'^{--}$ under
the standard model.  Including all these Higgs multiplets, we obtain a
one-loop running of $\sin^2\theta_W$.  Therefore, the upper bound of the
SU(3)$_L$
breaking, $u$, can be computed from the equation $\sin^2\theta_W~(u) = 1/4 $.
Since the result is very sensitive to the value of $\sin^2\theta_W$ at
$M_Z$, we plot in Fig.~1 the breaking scale $u$ as a
function of $\sin^2 \theta_W~(M_Z)$ for $\alpha^{-1}_{\rm em} = 127.9$
\cite{lang&luo}
in the $\overline{MS}$ scheme.  In particular,
for $\sin^2 \theta_W(M_Z) = 0.2333$
\cite{lang&luo}, we obtain that the breaking scale
is less than 1.7~TeV.

The breaking of the SM to U(1)$_{\rm em}$ can be achieved by $\langle \Delta^0
\rangle = v/\sqrt{2}, \; \langle \Delta '^0 \rangle = v'/\sqrt{2}$ and $\langle
\eta^0 \rangle = w/\sqrt{2}$, where $\langle \eta '^0 \rangle = 0$ is assumed
for lepton number conservation.  The charged gauge bosons
\begin{mathletters}
\begin{eqnarray}
W^+ &=& (W^1 - iW^2)/\sqrt{2} \ ,\phantom{AAAAA}  \\
Y^+ &=& (W^6 - iW^7)/\sqrt{2} \ ,\phantom{AAAA}  \\
\noalignand
\vspace{-1.5cm}
Y^{++} &=& (W^4 - iW^5)/\sqrt{2} \ ,\phantom{AAAAA}
\end{eqnarray}
\end{mathletters}
acquire masses
\begin{mathletters}
\begin{eqnarray}
M^2_W &=& \textstyle\frac{1}{4} \: g^2 (v^2 \!+\!v'^2 \!+\! w^2) \ , \\
M^2_{Y^+} &=& \textstyle\frac{1}{4} \: g^2(u^2 \!+\!v^2 \!+\! w^2) \ , \\
\noalignand
M^2_{Y^{++}} &=& \textstyle\frac{1}{4} \: g^2 (u^2
         \!+ v'^2 \!+ 4 w^2) \ ,\phantom{a}
\end{eqnarray}
\end{mathletters}
respectively.  For $v'^2 = w^2 = 0$, we have an approximate mass
relation, $M^2_{Y^{\pm}} = M^2_{Y^{\pm\pm}} + M^2_W$.  Therefore, we
would expect $Y^{\pm}$ to be heavier than $Y^{\pm\pm}$.

The mass-squared matrix for the neutral gauge bosons $\{ W^3, \:
W^8, \: V \}$ is given by
\begin{eqnarray}
\left[ \matrix{\frac{1}{4} g^2(v^2\!+\!v'^2\!+\!w^2) & - \frac{1}{4\sqrt{3}}
g^2
(v^2\!-\!v'^2\!+\!w^2) & - \frac{1}{2\sqrt{6}} gg_X\,v'^2\cr
-\frac{1}{4\sqrt{3}} g^2(v^2\!-\!v'^2\!+\!w^2) & \frac{1}{12} g^2(4u^2
\!+\!v^2\!+\!v'^2\!+\!w^2) & -\frac{1}{6\sqrt{2}} g g_X (2u^2\!+\!v'^2)\cr
-\frac{1}{2\sqrt{6}} gg_X\,v'^2 & - \frac{1}{6\sqrt{2}} gg_X(2u^2+v'^2) &
\frac{1}{6} g^2_X (u^2\!+\!v'^2)} \right] \ .
\end{eqnarray}
We can easily identify the photon field $\gamma$
as well as the massive bosons $Z$ and $Z'$
\begin{mathletters}
\begin{eqnarray}
\gamma &=& + \sin \theta_W \, W^3 + \cos \theta_W \left( \sqrt{3} \tan \theta_W
         \, W^8 + \sqrt{1\!-\!3 \tan^2\theta_W} \, V \right) \ , \\
Z &=& + \cos \theta_W \, W^3 - \sin \theta_W \left( \sqrt{3} \tan \theta_W \,
       W^8 + \sqrt{1\!-\!3 \tan^2 \theta_W} \, V \right) \ ,  \\
\noalignand
Z' &=& - \sqrt{1\!-\!3 \tan^2\theta_W} \, W^8 + \sqrt{3} \tan
      \theta_W \, V \ ,
\end{eqnarray}
\end{mathletters}
where the mass-squared matrix for $\{Z, \:  Z'\}$ is given by
\begin{eqnarray}
{\cal M}^2 = \pmatrix {M^2_Z & M^2_{ZZ'}\cr
M^2_{ZZ'} & M^2_{Z'}}
\end{eqnarray}
with
\begin{mathletters}
\begin{eqnarray}
M^2_Z &=& \frac{1}{4} \: \frac{g^2}{\cos^2\theta_W} \: (v^2\!+\!v'^2\!+\!w^2) \
          ,  \\
M^2_{Z'} &=& \frac{1}{3} \: g^2 \left[
              \frac{\cos^2\theta_W}{1\!-\!4 \sin^2\theta_W} \: u^2
           \: + \:  \frac{1\!-\!4 \sin^2\theta_W}{4\!\cos^2\theta_W}\:
                                      (v^2\!+\!v'^2\!+\!w^2) \hfill
\right. \nonumber \\
          &&\qquad\left.
          \: +\: \frac{3\!\sin^2\theta_W}{1\!-\!4 \sin^2\theta_W}\: v'^2
          \right] \ ,  \\
M^2_{ZZ'} &=& \frac{1}{4\sqrt{3}} \, g^2 \left[
         \frac{\sqrt{1\!-\!4 \sin^2\theta_W}}{\cos^2\theta_W} (v^2\!+\!w^2) -
       \left( \frac{1\!+\!4\sin^2\theta_W}{1\!-\!4\sin^2\theta_W} \right)
       v'^2 \right]  .
\end{eqnarray}
\end{mathletters}
The mass eigenstate are
\begin{mathletters}
\begin{eqnarray}
Z_1 &=& \cos\theta\ Z - \sin\theta\ Z' \ ,  \\
\noalignand
Z_2 &=& \sin\theta\ Z + \cos\theta\ Z' \ ,
\end{eqnarray}
\end{mathletters}
where the mixing angle is given by
\begin{equation}
\tan^2\theta=\frac{M_Z^2-M_{Z_1}^2}{M_{Z_2}^2-M_Z^2} \ .
\end{equation}
with $M_{Z_1}^2$ and $M_{Z_2}^2$ being the masses for $Z_1$ and $Z_2$.  Here,
$Z_1$ corresponds to the standard model neutral gauge boson and $Z_2$
corresponds to the additional neutral gauge boson.

Since $1\!-\!4 \sin^2\theta_W \simeq 0.06$ and $v'^2 \ll u^2$, we can
conclude that $M^2_{ZZ'} \ll M^2_{Z'}$.  Hence, we obtain
\begin{mathletters}
\begin{eqnarray}
M^2_{Z_1} &=& M^2_Z\: \left(1\!-\!\frac{M^4_{ZZ'}}{M^2_Z M^2_{Z'}}\right)\ ,\\
M^2_{Z_2} &=& \frac{1}{3} \: g^2  \, \frac{\cos^2\theta_W}
                               {1\!-\!4 \sin^2\theta_W} \: u^2 \ ,\\
\noalignand
\theta &=& \frac{M^2_{ZZ'}}{M^2_{Z'}}\ .
\end{eqnarray}
\end{mathletters}
In particular, for $v'=0$, the mass of $Z_1$ is shifted by a factor of
$\displaystyle{1 - \frac{1}{3}(1\!-\!4 \sin^2\theta_W)
\frac{M^2_Z}{M^2_{Z'}}}$ which is naturally negligible.

{}From the symmetry breaking hierarchy, $\eta > v, \, v', \, w$, we obtain the
lower mass bound of $Z_2$
\begin{eqnarray}
M_{Z_2} &\displaystyle\mathop{>}_{\sim}& \sqrt{\frac{4}{3}}\:
           \frac{\cos^2\theta_W (M_{Z_2})}{\sqrt{1\!-\!4
\sin^2\theta_W(M_{Z_2})}}
              M_{Z_1} \,\nonumber  \\
&\displaystyle\mathop{>}_{\sim}& 400 \; {\rm GeV} \ .
\end{eqnarray}
In many extensions of the SM, such as SO(10), E$_6$, L-R models, {\it etc}.,
the masses of the additional neutral gauge bosons are unconstrained in
general.  From $Z$-$Z'$ mixing
\cite{precision}, the lower limits are typically from
200~GeV to 1000~GeV, depending on the models.  They can also be as heavy as the
unification scale.  This model, on the other hand, predicts $M_{Z'}$ to be
within 400~GeV and 1.7~TeV.  In addition, the masses of the new charged gauge
bosons $Y^+$ and $Y^{++}$ are expected to be
\begin{eqnarray}
M_{Y^+} \simeq M_{Y^{++}} \ = \ M_Y \
   \simeq \sqrt{\frac{3}{4}} \: \frac{\sqrt{1\!-\!4
   \sin^2\theta_W}}{\cos \theta_W} \: M_{Z_2} \ .
\end{eqnarray}
which is depicted numerically in Fig.~2.  We find that $M_Y$ is
always less than $0.5 ~ M_{Z_2}$. Therefore, we expect that
the decays $Z' \to Y^{++} Y^{--}$ and $Y^{\pm\pm} \to 2 \ell^\pm \;
(\ell \!=\! e,\,\mu,\,\tau)$ are allowed, leading to spectacular signatures
in the future colliders. From the collider experiments
\cite{fram&ng} and muon decay \cite{belt&etal},
$M_{Y^{++}}$ and $M_{Y^+}$ are greater than 210~GeV (95\% C.L.)
and 270~GeV (90\% C.L.) respectively.  From Fig.~2, we otain a limit,
$M_{Z_2} > 1.3$~TeV, for $M_Y > 270~$ GeV.

\vspace{0.3cm}

\section{FERMION MASSES}

The Yukawa interactions corresponding to the scalar multiplets $\Phi, \:
\Delta, \: \Delta '$ and $\eta$ are given as follows
\begin{mathletters}
\begin{eqnarray}
- {\cal L}(\Phi) &=& h^{1,2}_{D,S} \, Q_{1,2} (D^c,S^c) \Phi^\ast + h^3_T \,
Q_3
\, T^c\, \Phi \ ,  \\
- {\cal L}(\Delta) &=& h^3_t \, Q_3 t^c \Delta + h^{1,2}_{d,s,b} \, Q_{1,2}
(d^c, s^c, b^c) \Delta^\ast + h_e^{ij} \, \psi_i \psi_j \, \Delta^\ast \ ,
\\
- {\cal L} (\Delta ') &=& h^3_b \, Q_3 b^c \Delta ' + h^{1,2}_{u,c,t} \,
Q_{1,2} (u^c, c^c, t^c) \Delta'^\ast \\
\noalignand
- {\cal L}(\eta) &=& y^{ij}_e \, \psi_i\psi_j \eta \ ,
\end{eqnarray}
\end{mathletters}
where $h^{ij}_e$ and $y^{ij}_e$ are antisymmetric and symmetric matrices in
flavor space.

As SU(3)$_L \times {\rm U}(1)_X$ breaks down to SU(2)$_L \times {\rm U}(1)_Y$,
$D, \, S,\,T$ acquire masses which are expected to be less than 10~TeV.  As
$\Delta, \, \Delta '$ and $\eta$ break the SU(2)$_L \times {\rm U}(1)_Y$ to
U(1)$_{\rm em}$, all the usual fermions acquire masses.  We have redefined the
SU(3)$_L$ singlets in such a way that $m_t =
\textstyle\frac{1}{\sqrt{2}} \, h^3_t\,v$ and $m_b =
\textstyle\frac{1}{\sqrt{2}} \, h^3_b \, v'$.  Therefore, we expect that
$v' \ll v$ as we assumed in the previous section.  The mass matrix linking
the left-handed to right-handed quarks is given by
\begin{eqnarray}
\frac{1}{\sqrt{2}} \pmatrix {h^1_d v & h^2_d v & 0\cr
h^1_s v & h^2_s v & 0\cr h^1_b v & h^2_b v & \sqrt{2} m_b\cr} \ .
\end{eqnarray}
The mass matrix of the up-sector has the same form.  It would be natural
to assume that $m_b$ is much bigger than the other elements after the
redefinition.  Therefore, the mixing hierarchy will
be $D^L_{sb} = m_s/m_b$ and $D^L_{db} = m_d/m_b$
for the left-handed sector, whereas $D^R_{sb} = (m_s^2/m^2_b)$ and
$D^R_{db} = (m^2_d/m^2_b)$ for the right-handed sector.  Hence we
obtain the CKM matrix elements $V_{cb} \simeq m_s/m_b$ and
$V_{ub} \simeq m_d/m_b + D^L_{uc} (m_s/m_b) \simeq$
$D^L_{uc} (m_s/m_b).$  Thus
$V_{ub}/V_{cb} \simeq 0.1$
\cite{partdata}, implies $D^L_{uc} \simeq 0.1$.

For the lepton sector, the charged lepton mass matrix is $h^{ij}_e \,
v/\sqrt{2} + y^{ij}_e \, w/\sqrt{2}$.  Without the $\eta$, the matrix is
symmetric, yielding a unacceptable relationship, namely $m_\mu =
m_\tau$.  In addition, we have assumed $\langle \eta^{0 '} \rangle = 0$
so that neutrinos remain massless and there will be residual lepton number
conservation.  In general, if $\langle \eta^{0 '} \rangle \neq 0$,
heavy SU(3)$_L\times {\rm U}(1)_X$ singlet  neutrinos are required for
a see-saw mechanism in
order to obtain realistic masses for the light neutrinos.  Nevertheless,
assuming $\langle \eta^{0'} \rangle = 0$ will not affect our discussion in this
paper.

\vspace{0.3cm}

\section{FLAVOR-CHANGING NEUTRAL CURRENT PROCESSES}

In this model, the interactions of $Z'$ discriminate among quark generations.
Since $M_{Z'}$ is expected to be smaller than 1.7~TeV, flavor-changing neutral
current processes induced by $Z'$ would be important tests for this model.  As
explained in Sec.~II, the choice of an anti-triplet quark multiplet is
arbitrary; here we first choose the third generation.  To define the convention
properly, we explicitly write out all the fermions and neutral gauge bosons
$(A,\,Z$ and $Z')$ as follows:
\begin{eqnarray}
{\cal L}(A) =  Q_f \:e  \: A^\mu \, \overline{f} \, \gamma_\mu \,
              f \ ,\phantom{AAAAAAAAAA}
\end{eqnarray}
\begin{eqnarray}
{\cal L}(Z) = \frac{g}{\cos \theta_W} \: Z^\mu \, \overline{f} \, \gamma_\mu
    (g_V(f) + g_A (f) \gamma_5) f \ ,
\end{eqnarray}
with $g_V (f) = \frac{1}{2} \, T_f - Q_f \, \sin^2\theta_W$ and
$g_A(f) = - \frac{1}{2} \, T_f$, and
\begin{eqnarray}
{\cal L}(Z') = \frac{g}{\cos \theta_W} \: Z'^\mu \, \overline{f} \, \gamma_\mu
(a_f\!+\!b_f \, \gamma_5) f \ ,\phantom{AAAA}
\end{eqnarray}
with
\begin{mathletters}
\begin{eqnarray}
a_\nu &=& - b_\nu = \frac{1}{4\sqrt{3}} \: \sqrt{1\!-\!4\sin^2\theta_W} \ , \\
a_{e,\mu ,\tau} &=& 3b_{e,\mu,\tau} = 3a_\nu \ ,  \\
a_{u,c} &=& \frac{1}{4\sqrt{3}} \:
          \frac{-1\!+\!6 \sin^2\theta_W}{\sqrt{1\!-\!4 \sin^2\theta_W}}\,,
\quad b_{u,c} = \frac{1}{4\sqrt{3}} \: \frac{1\!+\!2 \sin^2\theta_W}
                   {\sqrt{1\!-\!4 \sin^2\theta_W}} \ ,  \\
a_t &=& \frac{1}{4\sqrt{3}} \: \frac{1\!+\!4 \sin^2\theta_W}
                                    {\sqrt{1\!-\!4 \sin^2\theta_W}} \,,
\quad b_t = - \frac{1}{4\sqrt{3}} \: \sqrt{1\!-\!4 \sin^2\theta_W} \ ,  \\
a_{d,s} &=& -\frac{1}{4\sqrt{3}} \: \frac{1}{\sqrt{1\!-\!4 \sin^2\theta_W}} \,
,
\quad b_{d,s} = \frac{1}{4\sqrt{3}} \: \sqrt{1\!-\! 4\sin^2\theta_W} \ ,  \\
a_b &=& \frac{1}{4\sqrt{3}} \: \frac{1\!-\!2 \sin^2\theta_W}
                            {\sqrt{1\!-\!4 \sin^2\theta_W}} \, ,
\quad b_b =-\frac{1}{4\sqrt{3}} \: \frac{ 1\!+\!2 \sin^2\theta_W}
                              {\sqrt{1\!-\!4 \sin^2\theta_W}} \ , \\
a_{D,S} &=& \frac{1}{2\sqrt{3}} \: \frac{1\!-\!9 \sin^2\theta_W}
                        {\sqrt{1\!-\!4 \sin^2\theta_W}} \, ,
\quad b_{D,S} = - \frac{1}{2\sqrt{3}} \: \frac{1\!-\!\sin^2\theta_W}
                                     {\sqrt{1\!-\! 4\sin^2\theta_W}} \ ,  \\
a_T &=& - \frac{1}{2\sqrt{3}} \: \frac{1\!-\! 11\sin^2\theta_W}
                                   {\sqrt{1\!-\!4 \sin^2\theta_W}} \, ,
\quad b_T = \frac{1}{2\sqrt{3}} \: \frac{1\!-\! \sin^2\theta_W}
                                      {\sqrt{1\!-\!4 \sin^2\theta_W}} \ ,
\end{eqnarray}
\end{mathletters}
where $Q_f = -1, \, \frac{2}{3}, \, -\frac{1}{3}, \, -\frac{4}{3}$
and
$\frac{5}{3}$ for $f = (e,\mu,\tau), \: (u,c,t), \: (d,s,b), \: (D,S)$ and $T$,
respectively.  The weak
isospin for fermion $f, \; T_f$, is defined as $\frac{1}{2}, \: -\frac{1}{2}$
and 0 for $(\nu_e,\nu_\mu,\nu_\tau,u,c,t)$, $(e,\mu,\tau,d,s,b)$
and $(D,S,T)$, respectively.  Thus the couplings of
$D, \, S$ and $T$ to the $Z$ boson are vector-like.  Since the third generation
transforms differently, their couplings to $Z'$ differ from those
for the first and second
generations, leading to the flavor-changing neutral currents (FCNCs).  In
particular, the FCNC in the down sector is given by
\begin{eqnarray}
{\cal L}_{\rm FCNC} = \frac{g}{\cos \theta_W} \: [-\ \sin\theta Z_1 +
\cos\theta Z_2]
\left\{ \overline{b}' \, \gamma_\mu \, \delta_L \left( \frac{1\!-\!\gamma_5}{2}
\right) b \right\} \ ,
\end{eqnarray}
where
\begin{eqnarray}
\delta_L = (a_b\!-\!a_d) + (b_b \!-\!b_d) =
\frac{1\!-\!\sin^2\theta_W}{\sqrt{3} \,
\sqrt{1\!-\!4 \sin^2\theta_W}} \ .
\end{eqnarray}
There is no FCNC for the right-handed currents as the right-handed fermions
transform identically.  The $B^0$--$\overline{B}{}^0$ mixing will be calculated
to be
\begin{eqnarray}
\Delta {M}_{B^0_d} \simeq \frac{4 \pi\alpha}{3 \sin^2\theta_W \cos^2\theta_W}
\:
                \left(\frac{m_d}{m_b} \right)^2 \, \delta^2_L
    \left[ \frac{\cos^2\theta}{M^2_{Z_1}} + \frac{\sin^2\theta}{M^2_{Z_2}}
     \right] \,
      B_B f^2_B \, {M}_B \ ,
\end{eqnarray}
where $f_B$ and $B_B$ are the decay constant and bag factor of a $B$-meson.
Taking $\sin^2\theta_W = 0.2333, \; \sqrt{B_B} f_B = 160$~MeV,
$m_b = M_B = 5.3$~GeV, $m_d > 5$~MeV and $\Delta  M_B < 3\times 10^{-13}$~GeV,
we obtain $\sin\theta < 0.2$ and $M_{Z'} > 180$~GeV.  Because of the
suppression factor $(m_d/m_b)^2$, the $Z'$ mass
and the mixing angle $\sin\theta$ are not stringently constrained.
For $K^0$--$\overline{K}{}^0$ mixing, there will be an additional
suppression factor $(m_s/m_b)^2$ which leads to a negligible contribution.

$b$~--~$s$ transitions can also be induced by the Yukawa interactions
\begin{eqnarray}
\left[ \sqrt{2} \, \frac{m_b}{v} \: \Delta^0 - \sqrt{2} \:
                                   \frac{ m_b}{v'} \: \Delta '^{0} \right]
      D^L_{sb} \  \overline{b} \, \left( \frac{1\!-\!\gamma_5}{2} \right) s \ .
\end{eqnarray}
Assuming $v' > 10$~GeV and $m_{\Delta^{0'}} > 250$~GeV the contribution to
$\Delta  M_{B^0_d}$ is less than $3 \times 10^{-13}$~GeV. The contribution
from the first term is even smaller as $v \ \simeq \ 250$ GeV.

On the other hand, if the first or  second generation is
chosen to be an anti-triplet
of SU(3)$_L$, the $K^0$--$\overline{K}{}^0$ mixing is unsuppressed.
As a result,
$M_{Z'}$ has to be greater than 40~TeV
\cite{pis&ple}.  This is in contradiction
with the analyses in Sec.~III.  Therefore, in order that this model be viable,
the third  generation should be chosen to be the SU(3)$_L$
anti-triplet.  In
addition, due to the mixing hierarchy, the new contributions to
$B^0_s$-$\overline{B}^0_s$ mixing would be important.

\vspace{0.3cm}

\section{NEUTRINO ISOSCALAR SCATTERING}

Models with additional neutral gauge bosons, such as E$_6$ and L-R models, have
been intensively investigated
\cite{precision}.  In particular, the mass bounds at 90\% C.L.
for $Z_2$ from neutrino-hadron scattering for E$_6$ ($\chi$-model) and
L-R models
are 555~GeV and 795~GeV
\cite{aquilo&etal}, respectively.  The corresponding mixing angles
are bound to be less than $6 \times 10^{-3}$ and $5 \times 10^{-3}$.  Here, the
coupling strength of $Z'$ to quarks in this model is stronger than that of
leptons; neutrino-quark scattering would be the most important process among
the precision measurements.

In the low-energy limit, the four-fermion interactions are given by
\begin{eqnarray}
{\cal L} = - \frac{4 G_F}{\sqrt{2} }\:
    (J_\mu J^\mu - 2\delta\xi J_\mu J'^\mu + \xi J'_\mu J'^\mu) \ ,
\end{eqnarray}
where
\begin{eqnarray}
J_\mu = \overline{f} \, \gamma_\mu (g^f_V + g^f_A  \, \gamma_5)  f \ , \qquad
J'_\mu = \overline{f} \, \gamma_\mu (a^f + b^f \, \gamma_5) f \nonumber \\
\end{eqnarray}
and
\begin{eqnarray}
\delta = \frac{M^2_{ZZ'}}{M^2_{Z}}  , \quad \xi = \frac{M^2_{Z}}{M^2_{Z'}},
         \quad \overline\rho = 1\!-\!{\delta^2}{\xi} \ .
\end{eqnarray}
The above expressions, which can be obtained by taking the inverse of
Eq.~(3.9),  are exact without any approximation.
Hence, the ratio of neutral to charged current cross sections, ${\cal R}_\nu$,
given by
\begin{eqnarray}
{\cal R}_\nu = \left[ (\epsilon^{u}_L)^2 \!+\! (\epsilon^{d}_L)^2 \right]
         + \left[(\epsilon^{u}_R)^2 \!+\! (\epsilon^{d}_R)^2 \right]  r \ ,
\end{eqnarray}
where
\begin{mathletters}
\begin{eqnarray}
\epsilon^u_L &=& \frac{1}{\overline\rho}
      \left[ \rho_{\nu N} \left( \frac{1}{2} \!-\!
        \frac{2}{3}\, \kappa_{\nu N} \, \sin^2\theta_W \!+\!
                                                   \lambda^u_L \right)
          - \delta \; \xi \, \frac{\sqrt{1\!-\!4 \sin^2\theta_W}}{4\sqrt{3}} \,
         \left( \frac{1}{1\!-\!4 \sin^2\theta_W} + \frac{4}{3}\,
               \sin^2\theta_W \right)
       \right. \nonumber \\
   &&\quad\left. - \frac{1}{12} \, \xi (1\!-\!4 \sin^2\theta_W)
        \vphantom{\frac{sqrt{\sin^2\theta_W}}{\sqrt{3}}}
      \right] \ ,  \\
\epsilon^u_R &=& \frac{1}{\overline\rho}
       \left[ \rho_{\nu N} \left( -\frac{2}{3} \, \kappa_{\nu N} \,
            \sin^2\theta_W \!+\! \lambda^u_R \right) - \delta \: \xi \,
           \frac{\sqrt{1\!-\!4 \sin^2\theta_W}}{4\sqrt{3}}
         \, \left( \frac{8 \sin^2\theta_W}{1\!-\!4 \sin^2\theta_W} -
                 \frac{4}{3}\, \sin^2\theta_W \right)
       \right. \nonumber \\
&&\quad\left. + \frac{1}{3} \, \xi \: \sin^2\theta_W)
        \vphantom{\frac{sqrt{\sin^2\theta_W}}{\sqrt{3}}}
   \right] \ ,  \\
\epsilon^d_L &=& \frac{1}{\overline\rho}
      \left[ \rho_{\nu N} \left( -\frac{1}{2}\!+\! \frac{1}{3}\,
           \kappa_{\nu N} \, \sin^2\theta_W \!+\! \lambda^d_L \right) +
             \delta \ \xi \, \frac{\sqrt{1\!-\!4 \sin^2\theta_W}}{4\sqrt{3}} \,
                      \left( 2 \!+\! \frac{1}{1\!-\!4 \sin^2\theta_W} -
                  \frac{2}{3}\, \sin^2\theta_W \right)
      \right. \nonumber \\
&&\quad\left. - \frac{1}{12} \, \xi (1 \!-\! 2 \sin^2\theta_W)
    \vphantom{\frac{sqrt{\sin^2\theta_W}}{\sqrt{3}}} \right] \ ,  \\
\epsilon^d_R &=& \frac{1}{\overline\rho}
          \left[ \rho_{\nu N} \left( \frac{1}{3} \, \kappa_{\nu N} \,
                 \sin^2\theta_W \!+\! \lambda^d_R \right) +
             \delta \ \xi \, \frac{\sqrt{1\!-\!4 \sin^2\theta_W}}{4\sqrt{3}}
                \, \left( \frac{4 \sin^2\theta_W}{1\!-\!4 \sin^2\theta_W} -
                      \frac{2}{3}\, \sin^2\theta_W \right)
          \right. \nonumber \\
&&\quad\left. - \frac{1}{6} \, \xi \: \sin^2\theta_W)
         \vphantom{\frac{sqrt{\sin^2\theta_W}}{\sqrt{3}}} \right]
\end{eqnarray}
\end{mathletters}
is expected to impose the most stringent constraint on the parameters
$\delta$ and $\xi$.  $r$ in
Eq.~(6.4) is the ratio of antineutrino to neutrino scattering cross sections.
$\lambda$'s,
$\rho_{\nu N} \!-\!1$ and $\kappa_{\nu N} \!-\!1$ are the electroweak radiative
corrections.  Assuming that they are dominated by the oblique corrections, we
then obtain the approximate expressions
$\rho_{\nu N} \simeq 1 + \displaystyle\frac{3G_F}{8\sqrt{2}\pi^2}\, m^2_t$
and $K_{\nu N} \simeq 1$ for the $\overline{MS}$ renormalization scheme
\cite{partdata,sirlin}.

Using the most precisely measured values for ${\cal R}_\nu$ obtained by the
CDHS
\cite{abramowicz} and CHARM \cite{allaby&etal} collaborations, we plot
\begin{eqnarray}
\chi^2 = \left( \frac{{\cal R}_\nu ({\rm CDHS}) - {\cal R}_V}
                      {\sigma({\rm CDHS})}
          \right)^2
      + \left( \frac{{\cal R}_\nu ({\rm CHARM}) - {\cal R}_V}
                      {\sigma({\rm CHARM})}
          \right)^2
\end{eqnarray}
for $\chi^2 = 0.5$ and $2$ in Fig.~3, where we take the global fit for the
top-quark mass  $m_t = 124$~GeV. In this paper, we do not
consider a comprehensive analysis of this
additional neutral gauge boson from all precision measurements.  Instead, we
have shown in Fig. 3 that the parameter $M^2_{Z'}$ ($\simeq M^2_{Z_2}$)
is not stringently constrained.
For example, for $|\delta| < 0.5, \; M_{Z'}$(or $M_{Z_2}$)
can be anywhere in the allowed region.
\section{CONCLUSION}

In this paper, we have considered an electroweak theory of
SU(3)$\times$U(1) which introduces three new quarks $D, S$ and $T$ with
charges $-4/3$, $-4/3$ and $5/3$.  All the lepton generations transform
identically under this gauge symmetry; whereas one of the quark
generations, which is an SU(3) anti-triplet, transforms differently
from the other two.  An SU(3) triplet Higgs scalar, $\Phi$,  is required to
breaking the symmetry into the standard model.  Two SU(3) triplets,
$\Delta$ and $\Delta'$, are responsible for breaking the standard model
as well as providing masses for the usual fermions.  To
obtain realistic lepton masses, a sextet, $\eta$, is needed.
{}From the flavor changing neutral current processes, the third generation
quark is chosen to be the SU(3) anti-triplet.

By matching the coupling constants at the
symmetry breaking scale, we find that $\sin^2\theta_W$ should be less
than 1/4, leading to a breaking scale under $1.7$~TeV for
$\sin^2\theta_W(M_{Z_1})=0.2333$.  From the symmetry breaking hierarchy,
namely $1.7$~TeV $ > u > v, v', w$,  the mass of the additional
neutral gauge boson $Z_2$ ranges from $400$ GeV to $1.7$ TeV. In addition,
the mass of the new charged gauge bosons $Y^{\pm}$ and $Y^{\pm\pm}$ is less
than a half of $M_{Z_2}$. Therefore the decay $Z_2 \rightarrow Y^{++} ~
Y^{--}$ with $Y^{\pm\pm} \rightarrow 2 ~ \ell ^{\pm}$ provides unique
signatures in the future colliders.

{}From the muon decay experiments, $M_Y$ is
found to be at least $270$~GeV at $90 \%$ confidence level.
Hence, we obtain narrow windows for $M_Y$ and $M_{Z_2}$,
$270$ GeV $<M_Y<330$ GeV and  $1.3$ TeV $<M_{Z_2}<1.7$ GeV.
Therefore, this model can be either discovered or ruled out
at the future colliders.

\acknowledgments
The author would like to thank James T. Liu for reading the manuscript.
The work was supported in part by the Natural Science and Engineering
Research Council of Canada.

\figure{the breaking scale $u$ as a function of $\sin^2\theta_W(M_{Z})$}
\figure{the new charged gauge boson mass $M_Y$ as a function of $M_{Z'}$
	for $\sin^2\theta_W (M_Z) = 0.2333 \pm 0.0016$}
\figure{contour plot of $\chi^2 = 0.5$(solid line) and $2$(dotted line)
        as a function of $\delta$ and $\xi$}

\end{document}